\documentclass[aps,prl,reprint,showpacs,superscriptaddress,fleqn]{revtex4-1}

\usepackage{graphicx}
\usepackage[utf8]{inputenc}
\usepackage[english]{babel}
\usepackage{color}
\usepackage{soul}

\newcommand{\LPS}[1]{\textcolor{black}{#1}}
\newcommand{\LPSv}[1]{\textcolor{black}{#1}}

\begin{document}



\title{\LPSv{Common universal behaviors of magnetic domain walls \\driven by spin-polarized electrical current and magnetic field}} 


\author{R. D\'iaz Pardo}
\affiliation{Laboratoire de Physique des Solides, Universit\'e Paris-Sud, Universit\'e Paris-Saclay, CNRS, UMR8502, 91405 Orsay, France.}
\author{N. Moisan}
\affiliation{Laboratoire de Physique des Solides, Universit\'e Paris-Sud, Universit\'e Paris-Saclay, CNRS, UMR8502, 91405 Orsay, France.}
\author{L. Albornoz}
\affiliation{Laboratoire de Physique des Solides, Universit\'e Paris-Sud, Universit\'e Paris-Saclay, CNRS, UMR8502, 91405 Orsay, France.}
\affiliation{Instituto de Nanociencia y Nanotecnología CNEA-CONICET, Centro Atómico Bariloche, Av. Bustillo 9500, R8402AGP, San Carlos de Bariloche, Argentina.}
\affiliation{Instituto Balseiro, Univ. Nac. Cuyo - CNEA, Av. Bustillo 9500, R8402AGP, S. C. de Bariloche, Rio Negro, Argentina.}

\author{A. Lema\^{\i}tre}
\affiliation{Centre de Nanosciences et de Nanotechnologies (C2N), CNRS, Univ. Paris-Sud, Universit\'e Paris-Saclay, 91120 Palaiseau, France.}
\author{J. Curiale}
\affiliation{Instituto de Nanociencia y Nanotecnología CNEA-CONICET, Centro Atómico Bariloche, Av. Bustillo 9500, R8402AGP, San Carlos de Bariloche, Argentina.}
\affiliation{Instituto Balseiro, Univ. Nac. Cuyo - CNEA, Av. Bustillo 9500, R8402AGP, S. C. de Bariloche, Rio Negro, Argentina.}
\author{V. Jeudy}
\email{vincent.jeudy@u-psud.fr}
\affiliation{Laboratoire de Physique des Solides, Universit\'e Paris-Sud, Universit\'e Paris-Saclay, CNRS, UMR8502, 91405 Orsay, France.}


\date{\today}
\begin{abstract}

We explore universal behaviors of magnetic domain wall driven by \LPSv{the spin-transfer of an electrical current}, in a ferromagnetic (Ga,Mn)(As,P) thin film with perpendicular magnetic anisotropy. 
\LPSv{For a current direction transverse to domain wall, the dynamics of the thermally activated creep regime and the depinning transition are found to be compatible with a self-consistent universal description of magnetic field induced domain wall dynamics.
This common universal behavior, characteristic of the so-called quenched Edwards-Wilkinson universality class, is confirmed by a complementary and independent analysis of domain wall roughness.
However, the tilting of domain walls and the formation of facets is produced by the directionality of interaction with the current, which acts as a magnetic field only in the direction transverse to domain wall.} 

\end{abstract}

%

\maketitle
  

The displacement of small spin texture as magnetic domain walls (DWs) 
thanks to spin torque effects is at the basis of potential applications to magnetic memory storage~\cite{parkin_naturenano_2015_race_track}.
An important effort is dedicated to search for
magnetic materials~\cite{caretta_natnano_2018,hrabec_nanotech_2019} with large and well controlled DW velocities.
However, DWs are very sensitive to weak pinning defects~\cite{lemerle_PRL_1998_domainwall_creep,Moon_PRL_2013}, which strongly reduce their mobility and produce roughening and stochastic avalanche-like motion~\cite{ferrero_prl_2017_spatiotemporal_patterns,grassi_prb_2018}.  Therefore, 
it is particularly interesting to better understand the contribution of pinning to current induced DW dynamics. 

Magnetic domain walls~\cite{lemerle_PRL_1998_domainwall_creep,Yamanouchi_science_2007,ferrero_prl_2017_spatiotemporal_patterns,grassi_prb_2018,metaxas_PRL_07_depinning_thermal_rounding,Moon_PRL_2013,duttagupta_natphys_2015,jeudy_PRL_2016_energy_barrier,diaz_PRB_2017_depinning}  
present surprising universal critical behaviors, encountered in a wide variety of moving interfaces such as the reaction front propagation in disordered flows~\cite{atis_prl_2015}, growing bacterial colonies~\cite{Bonachela_bacteria_colonies_2011}, wetting \cite{moulinet_EPJE_2002}, motion of ferroelectric domain walls~\cite{tybell2002}, to name a few. 
The interfaces are rough with self-affine width growing as  $L^{\zeta}$, where $L$ the distance between two points of the interface and $\zeta$ the roughness exponent.
Moreover, a depinning driving force $f_d$ separates the so-called creep ($f <f_d$) and depinning ($f \gtrsim f_d$) regimes. 
In the creep regime, the velocity varies as an Arrhenius law $v \sim e^{-\Delta E/k_BT}$~\cite{lemerle_PRL_1998_domainwall_creep,chauve_2000, jeudy_PRL_2016_energy_barrier}, where $k_BT$ is the thermal fluctuation energy. $\Delta E$ is the effective pinning energy barrier height, which follows a universal power law variation with the driving force $\Delta E \sim f^{-\mu}$, where $\mu$ is the creep exponent. 
In the depinning regime~\cite{LeDoussal2009,bustingorry_PRE_12_thermal_rounding,diaz_PRB_2017_depinning}, the effective pinning barriers are collapsed. The velocity presents power law variations with drive $f$ and temperature $T$: $v \sim (f-f_d)^\beta$ and $v(f_d) \sim T^\psi$, where $\beta$ and $\psi$ are the depinning and thermal rounding exponents, respectively.

Universal behaviors have been extensively investigated for DWs driven by magnetic field ($f \propto H$) in ferromagnetic ultrathin films. 
For a large variety of materials, the measured values of the creep ($\mu=1/4$)~\cite{lemerle_PRL_1998_domainwall_creep,metaxas_PRL_07_depinning_thermal_rounding,duttagupta_natphys_2015,jeudy_PRL_2016_energy_barrier} and roughness ($\zeta \approx 0.66$~\cite{lemerle_PRL_1998_domainwall_creep,metaxas_PRL_07_depinning_thermal_rounding,Moon_PRL_2013} and $\zeta \approx 1.25$~\cite{grassi_prb_2018} ) exponents,
are compatible with the prediction
for the motion of an elastic 1D line in a short-range weak pinning disorder, described by the so-called quenched Edwards Wilkinson (qEW) universality class with~\cite{rosso_prl_2001}, and without~\cite{chauve_2000,kolton_prb_2009_pathways} anharmonic contributions, respectively.
\LPSv{Moreover, it was recently shown~\cite{diaz_PRB_2017_depinning} that the depinning transition is compatible with the predictions for the qEW universality class ($\beta=0.25$ ~\cite{LeDoussal2009}, and $\psi=0.15$~\cite{bustingorry_PRE_12_thermal_rounding}).}

In contrast, the universal behaviors of DW motion induced by spin-polarized electric current are more contentious. To the best of our knowledge, the universality of the depinning transition has not yet been explored. For the creep motion, a compatibility with $\mu=1/4$ is suggested for DW driven by the conventional spin transfer torque (STT) in Pt/Co/Pt nanowires~\cite{lee_prl_2011} and by spin orbit torque (SOT), in ferrimagnets~\cite{caretta_natnano_2018}. 
\LPS{However, rather intriguing differences between current and magnetic field driven motion are also reported in the literature.}
Different values of the creep exponent were reported for other materials  ($\mu=0.33 \pm 0.06$ for (Ga,Mn)As~\cite{Yamanouchi_science_2007} and $\mu=0.39 \pm 0.06$ for Ta/CoFeB/MgO~\cite{duttagupta_natphys_2015}), which are difficult to interpret.
\LPS{The tilting and faceting~\cite{Moon_PRL_2013,Moon_natcomm_2018} of DWs, produced by the current} could suggest a compatibility with the so-called quenched Kardar-Parisi-Zhang (qKPZ) universality class~\cite{Moon_PRL_2013}. However, in the direction perpendicular to DW, the roughness is characterized by an exponent ($\zeta_j=0.69 \pm 0.04$) independent of DW tilting angle and compatible with the measurement obtained for field driven DW motion ($\zeta_H=0.68 \pm 0.04$), while a different value ($\zeta_j=0.99 \pm 0.01$) is obtained in the direction of current. 
Therefore, whether distinct or common universality classes describe the motion of DWs produced by current and magnetic field remains an open question.
\LPSv{In this letter, we show that the roughness and universal dynamics of spin-transfer-torque (STT) driven domain wall are particularly well described within the qEW universality class.}
	

\textit{Experimental techniques}. 
The experiments were performed with rectangles of a 4 nm thick (Ga,Mn)(As,P)/(Ga,Mn)As bilayer film patterned by lithography. The film was grown on a (001) GaAs/AlAs buffer ~\cite{niazi_apl_2013}. It has an effective perpendicular anisotropy and a Curie temperature ($T_c$) of 65~K. 
The sizes of rectangles were 133 $\times$ 210, 228 $\times$ 302, and 323 $\times$ 399 $\mu$m$^2$ (see supplemental material~\cite{supplemental} for the details). Two 40 $\mu$m wide gold electrodes (separated by 110, 204, and 300 $\mu$m, respectively) were deposited by evaporation parallel to the narrow sides of rectangles~\cite{supplemental}. \LPS{They were used to generate an homogeneous current density producing DW motion by STT.} 
The pulse amplitude varied between 0 and 11~$\,\mathrm{GA/m^2}$.
We verified that the Joule effect had a negligible contribution on DW dynamics \cite{supplemental}.
Perpendicular magnetic field pulses of adjustable amplitude (0-65~mT) were produced by a 75 turns small coil (diameter $\sim$ 1 mm) mounted on the sample.
The set-up was placed in an optical He-flow cryostat allowing to stabilize the temperature between 5.7~K and $T_c$. The DW displacement is observed using a magneto-optical Kerr microscope (resolution $\sim$ 1~$\mu m$). The DW velocity is defined as the ratio between the average displacement $\Delta x$ and the pulse duration $\Delta t$ \cite{supplemental}, which varies between 1~$\mu$s and 120~s.

\begin{figure}[ht]
	\includegraphics[width=8.7 cm,height=6.5 cm]{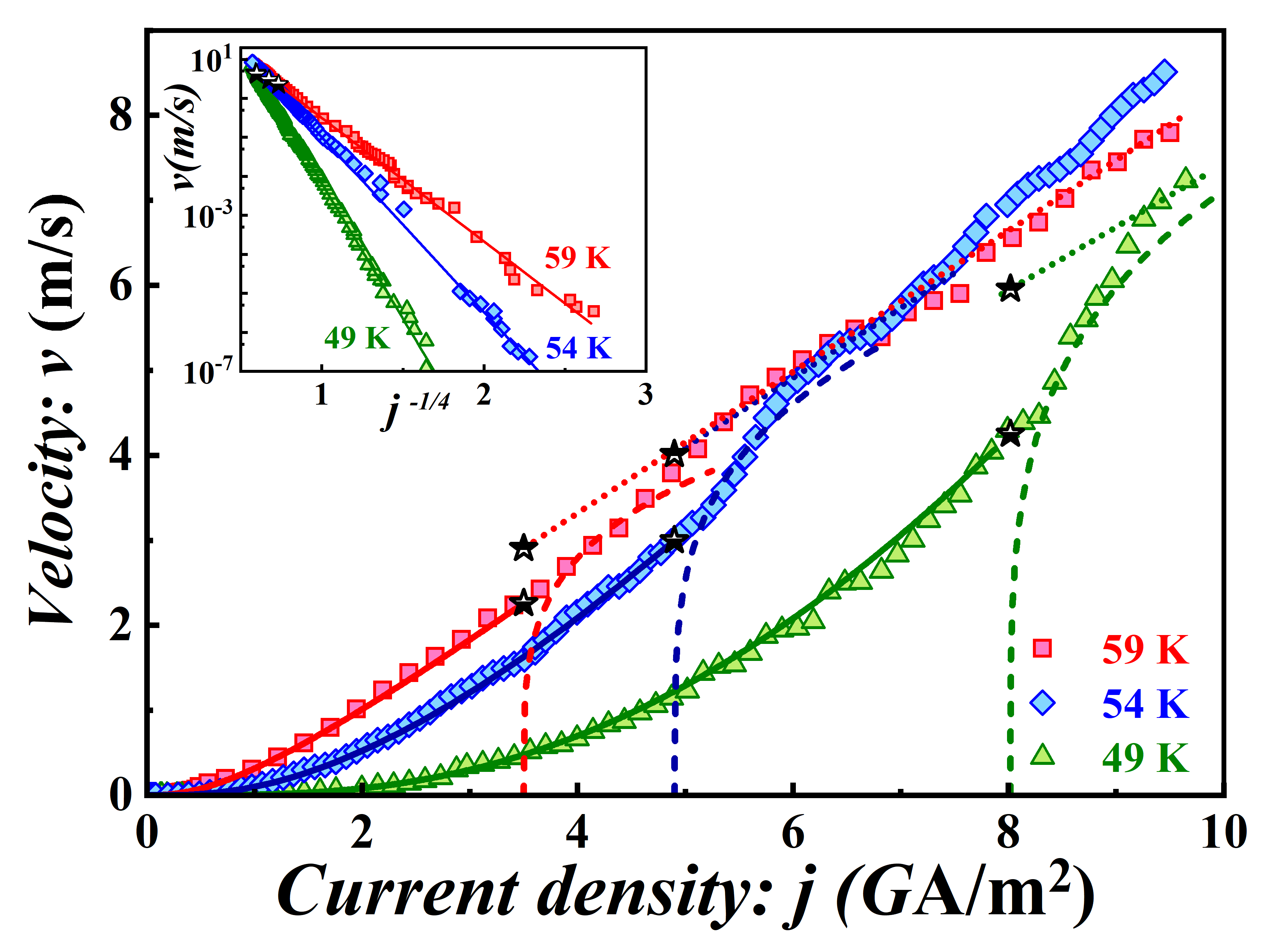}
	\caption{\textbf{Analysis of domain wall dynamics}. DW velocity versus current density $j$ for different temperatures. \LPS{The solid, dashed, and dotted lines are predictions for the creep, depinning and flow regimes, respectively, obtained from simultaneous fits of Eqs.~\ref{eq:velocity}. The lower and upper stars correspond to the velocities at the depinning threshold $v(j_d)$, and $v_T(j_d)$, respectively.} A sliding average over 5 points is used to smooth the curves. The unexpected large velocities measured for $T = 54$~K and $j>7.5$~$GA/m^2$ were not used for the fit. Inset: semi-log plot of the \LPS{same velocity curves} versus $j^{-1/4}$ highlighting the thermally activated creep regime and its fit. 
	} 
	\label{fig:3}
\end{figure}

\textit{DW dynamics and creep exponent}.
Let us start with investigations of current driven DW dynamics. In order to circumvent the variation of driving force with DW tilt, the velocity measurements \LPS{were all performed from almost flat DWs transverse to current ($\theta \approx 0$) (see \cite{supplemental}).}
The velocity curves are reported in Fig.~\ref{fig:3} and show similar features to those usually encountered in the literature for magnetic field driven DW dynamics~\cite{metaxas_PRL_07_depinning_thermal_rounding,jeudy_PRL_2016_energy_barrier,diaz_PRB_2017_depinning}. At low drive ($j<j_d$), the velocity follows a strong non-linear variation with drive and temperature, which characterizes the thermally activated creep regime (see the inset in Fig.~\ref{fig:3}). The curves present a change of curvature below and above the depinning threshold ($j=j_d$). The linear variation observed well above threshold corresponds to the flow regime. Here the exploration of DW dynamics was limited to the temperature range $T=$ 49-59~$K$, due to large thermal fluctuations impeding accurate displacement measurements above 59~$K$ and to an increase of the depinning threshold beyond experimental access below 49~$K$.

For more quantitative insights on \LPS{dynamical universal behaviors, the velocity curves were compared to the self-consistent description of creep, depinning and flow regimes developed for magnetic field driven DW motion~\cite{diaz_PRB_2017_depinning}:}
%
%
\begin{equation}
\label{eq:velocity}
v(j)=\left \{
\begin{array}{lr}
v(j_d)exp (-\frac{\Delta E}{k_B T}) & creep: j<j_d\\
\frac{v(j_d)}{x_0}(\frac{T_d}{T})^{\psi}(\frac{j-j_d}{j_d})^\beta & depinning: j \gtrsim j_d \\
v_T(j_d)\frac{j}{j_d}& linear flow: j \gg j_d, \\
\end{array}
\right.
\end{equation}
where $v(j_d)$ is the velocity at depinning threshold, and $v_T(j_d)=v(j_d)(T_d/T)^{\psi}$  the velocity that DW would reach at $j=j_d$ without pinning, i.e. within the flow regime. 
For the creep regime, the energy barrier height is given by $\Delta E=k_B T_d((j/j_d)^{-\mu}-1)$, where
$k_B T_d $ is the characteristic height of effective pinning barrier.
\LPS{For the depinning regime, Eq.~\ref{eq:velocity} is only valid over a limited range: 
it does not account for the effect of thermal fluctuations occurring just above the depinning threshold nor for the crossover to linear flow regime~\cite{diaz_PRB_2017_depinning}. 
The depinning exponents are $\beta =0.25$, and $\psi =0.15$, and the critical parameter $x_0 = 0.65$~\cite{diaz_PRB_2017_depinning}. Finally, the linear flow regime in Eq.~\ref{eq:velocity} is only reached at the end of crossover}.


\begin{table}[h]
	\centering
	\setlength\tabcolsep{1.5pt} 
	
	\begin{tabular}{|c||c|c|c||c|c|c|}
		\hline
		\hline  
		
		& \multicolumn{3}{c|}{Current}& \multicolumn{3}{c|}{Magnetic field} \\
		\hline
		
		\bf{$T$} & \bf{$ j_d$} & \bf{$ v(j_d)$} & \bf{$T_d$} & \bf{$H_d$} & \bf{$ v(H_d)$} & \bf{$T_d$}\\
		
		\hline
		49 & 7.9 (0.6)& 4.1 (0.5) & 465 (35) & 28.4 (0.4)& 7.5 (0.4) & 320 (15)\\
		54 & 4.9 (0.4)& 3.0 (0.4) & 375 (25) & 15.5 (0.4)& 4.0 (0.2) & 420 (20)\\
		59 & 3.5 (0.3)& 2.3 (0.2) & 315 (30) & 16.1 (0.5)& 4.2 (0.4) & 410 (20)\\
		\hline
		\hline		
	\end{tabular}
	\caption{\label{table:table1}  \textbf{Pinning parameters of DW dynamics.}   
	Fitting parameters of Eqs. \ref{eq:velocity}. {\it Units:} $T$ and $T_d$ are in Kelvin, \LPS{$v(j_d)$ and $v(H_d)$} in $\mathrm{m/s}$, $j_d$ in $\,\mathrm{GA/m^2}$ and, $H_d$ in $\mathrm{mT}$. 
	} 
	
\end{table}

\LPS{For each temperature, the fit of three regimes described by Eqs.~\ref{eq:velocity}  (see Fig.~\ref{fig:3}) relies only on three independent adjustable parameters ($j_d(T)$, $v(j_d(T))$, and $T_d$) and was performed simultaneously.
As the depinning threshold $j_d$ is not known {\it a priori}, the following procedure was used. The fit of creep regime was performed between the lowest measured velocity and increasing values of $v(j,T)$ on the velocity curve. $T_d(T)$ was taken as a free parameter, and $\mu =1/4$. \LPSv{The threshold velocity $v(j=j_d, T)$ was assumed to correspond to the best simultaneous fit of Eqs.~\ref{eq:velocity}.}  
As it can be observed in Fig. \ref{fig:3}, a good agreement is obtained with the data. The same analysis was performed for magnetic field driven DW motion~\cite{supplemental}. A comparison of the pinning parameters obtained in both cases is shown in Table \ref{table:table1}. The pretty close heights of effective pinning energy barrier ($k_BT_d$) reported for each temperature indicates that a similar weak pinning disorder controls both dynamics.}
\LPS{In order to estimate precisely the value of creep exponent $\mu$, it was set as a free shared parameter and we perform a global fit of the creep power law $ln v \sim j^{-\mu}$, for $j<j_d(T)$, and all the temperatures. The obtained values for field ($\mu_H=0.247 \pm 0.011$)~\cite{supplemental} and current ($\mu_j=0.259 \pm 0.004$) driven DW motion match within experimental error bars~\cite{comment}.
Therefore, a good agreement is obtained both for the creep exponent and self-consistent description of the creep and depinning universal dynamics. This is a strong evidence that DW motion induced by transverse current through the STT mechanism shares common universal creep and depinning behaviors of DW driven by magnetic field observed in thin magnetic films made of different materials~\cite{lemerle_PRL_1998_domainwall_creep,metaxas_PRL_07_depinning_thermal_rounding,duttagupta_natphys_2015,jeudy_PRL_2016_energy_barrier,diaz_PRB_2017_depinning,jeudy_PRB_2018_DW_pinning}.}

\textit{Roughness exponent}.
\begin{figure}
		\includegraphics[width=8.9cm]{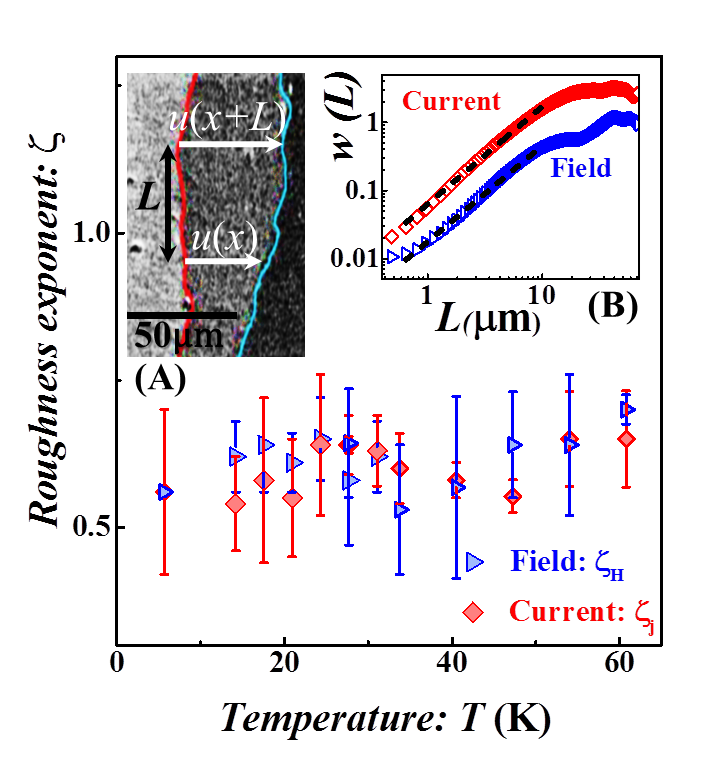}
	\caption{\textbf{Analysis of domain wall roughness}.  
		Roughness exponents $\zeta_j$ and $\zeta_H$ as a function of temperature, measured for $j<j_d$ and  $H<H_d$, respectively (see \cite{supplemental}). Inset (A): Definition of the domain wall displacements $u(x)$ and $u(x+L)$ used to calculate the displacement-displacement correlation function (see Eq. \ref{eq:correlation}). Inset (B): Typical correlation functions $w$ versus DW segment length $L$ in log-log scale, for current (diamonds) and field (triangles) driven DW ($T= 14.2 K$). The dashed lines are fit of $w\sim L^{2\zeta}$ used to determine the roughness exponents $\zeta_j$ and $\zeta_H$.	 	 
	} 
	\label{fig:2}
\end{figure}
\LPSv{As this finding do not confirm the analysis reported in Refs.~\cite{Yamanouchi_science_2007,duttagupta_natphys_2015}, we have investigated the DW roughness in the creep regime, in order to obtain a complementary and independent test of universal behavior.}
The DW self affinity was studied using the displacement-displacement correlation function~\cite{lemerle_PRL_1998_domainwall_creep}:
\begin{equation}
\label{eq:correlation}
w(L)=\sum_{x}[u(x+L)-u(x)]^2, 
\end{equation}
where $u(x)$ is the DW displacement measured parallel to current and $L$ the length of DW segment along the axis $x$ transverse to current (see the inset of Fig. \ref{fig:2} (A)). For a self-affine interface, the function $w(L)$ is expected to follow a power law variation $w(L)\sim L^{2\zeta}$, where $\zeta$ is the roughness exponent.
Typical variations of $w(L)$ versus $L$ obtained for field and current induced motion are compared in Fig.~\ref{fig:2} (B) in log-log scale. \LPSv{As it can be observed, both curves present a linear variation with similar slopes ($= 2\zeta$), between the microscope resolution ($\approx$ 1~$\mu m$) and $L=$10~$\mu m$. In order to get more statistics, the slopes were systematically determined }  
for successions of DW positions (see Ref.~\cite{supplemental}) and a temperature varying over one decade ($T=$4.5-59~$K$). 
The mean and standard deviation of the roughness exponent for current ($\zeta_j$) and field ($\zeta_H$) driven DW motion is reported in Fig. \ref{fig:2} as a function of temperature. As expected for universal critical exponents, the values of $\zeta_j$ and $\zeta_H$ do not vary significantly.
Their mean values ($\zeta_j=0.60 \pm 0.05$ and $\zeta_H=0.61 \pm 0.04$), calculated from all measurements, agree well within experimental error.
\LPS{Here, it is important to notice some differences and similarities with the results reported for Pt/Co/Pt in Ref. ~\onlinecite{Moon_PRL_2013}. For Ga/Mn/As, $\zeta_j$ presents no significant variation with DW tilting (see~\cite{supplemental} for details) and remains always significantly smaller than the value ($\zeta_j=0.99 \pm 0.01$) reported for DW displacements measured in the direction of current~\cite{holl_PRE_2019,comment_2}.
%
In contrast, the value $\zeta_j=0.60 \pm 0.05$ is compatible with the results ($\zeta_j=0.68 \pm 0.04$) of Ref. ~\onlinecite{Moon_PRL_2013} for a DW roughness analyzed in the direction ($\vec{n}$) normal to DW. The good agreement is also obtained with usual experimental results reported in the literature for field driven DW motion~\cite{lemerle_PRL_1998_domainwall_creep,lemerle_PRL_1998_domainwall_creep,metaxas_PRL_07_depinning_thermal_rounding,Moon_PRL_2013}, and with theoretical predictions ($\zeta=0.635 \pm 0.005$)~\cite{kolton_prb_2009_pathways,Rosso_PRE_2003} for the universality class of the qEW model with short range disorder and elasticity including an-harmonic correction.} 
\LPS{Therefore, both the analysis of DW dynamics and roughness lead to the same conclusion of common universal behaviors for current and field driven DW motion.}

\begin{figure}[ht]
	\centering
	\includegraphics[width=8.7cm]{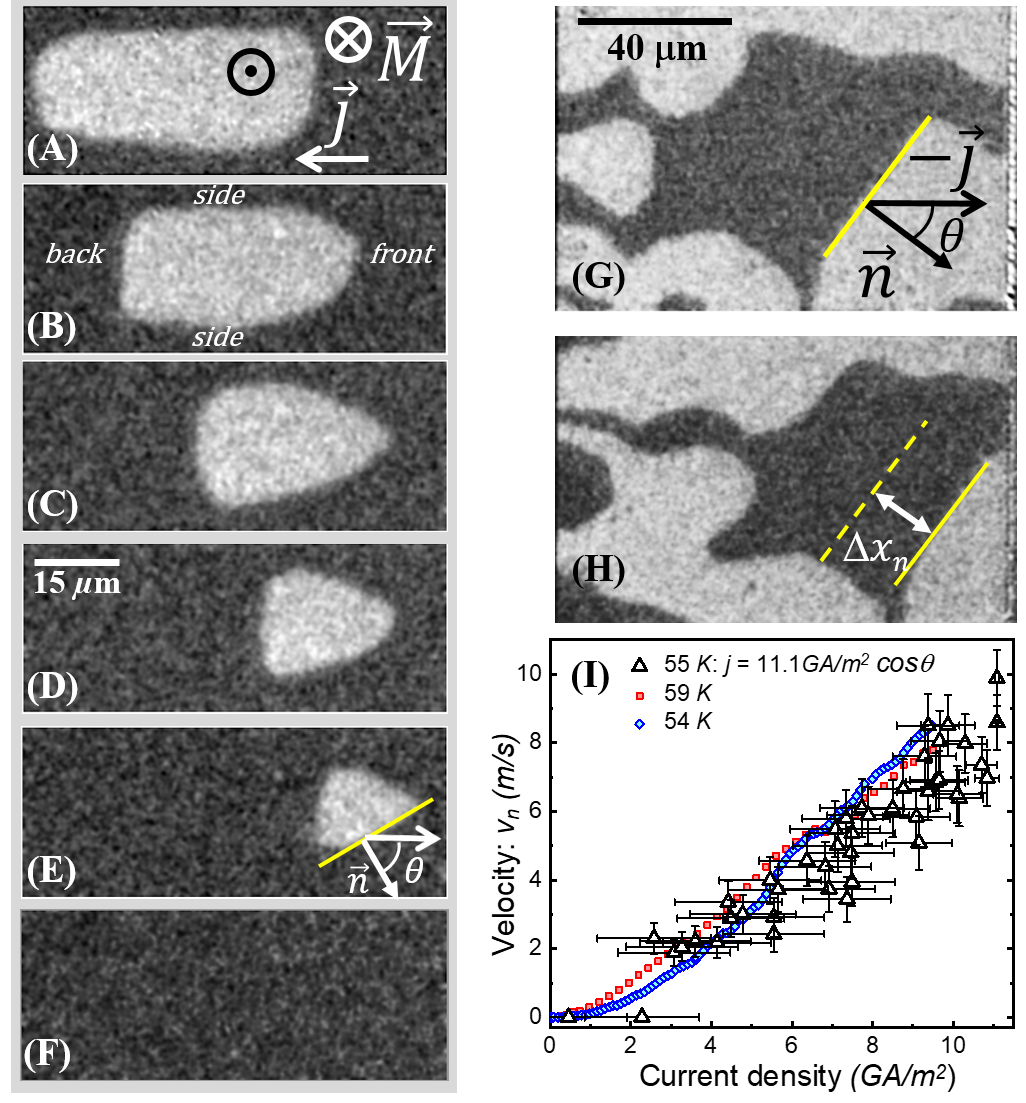}	
	\caption{\textbf{Current driven domain wall motion}  
		(A-F) Evolution of an almost rectangular magnetic domain submitted to pulses of current (duration \LPSv{$\Delta t=6 \mu s$} and amplitude $j=11\,\mathrm{GA/m^2}$) for $T = 45$~K. 
		The gray levels correspond to opposite directions of magnetization $\vec{M}$ perpendicular to the sample.
		%
		The back DW conserves its shape and velocity while the front DW becomes pointed, which reduces its velocity and leads to the collapse of the domain (see images E and F). 
		%
		(G-H) Two successive images 
		of domains ($\Delta t=1 \mu s$ and $j=11.1\,\mathrm{GA/m^2}$) showing, for $T = 55$~K, the displacement $\Delta x_n$ of a tilted DW (see the dashed and solid segments) along its normal direction $\vec{n}$. $\theta$ is the tilting angle between $\vec{n}$ and $-\vec{j}$. 
		\LPSv{(I) Curve of the velocity $v_n$ along $\vec{n}$ versus $\left|j\right| cos \theta$ obtained for a constant value of $j$ ($=11.1\,\mathrm{GA/m^2}$) for $T = 55$~K, and $\Delta t=1 \mu s$, and comparison with the velocity curves versus current density ($\theta =0$) taken from Fig.~\ref{fig:3}}. The error bars reflect the uncertainties of tilting angle and displacement. 
	} 
	\label{fig:5}
\end{figure}

{\it Origin of Domain wall faceting.}
\LPS{This common behavior, which is {\it a priori} difficult to reconcile with the faceting of DW produced by the current calls for investigation on the origin of DW faceting.}
%
First, we analyze the evolution of an initially almost rectangular domain subjected to a large current density ($j= 11 \mathrm{GA/m^2}$).  
As it can be observed in Fig.~\ref{fig:5}, the edges of the domain aligned along the current (i.e., $\vec{j} \perp \vec{n}$ 
, where $\vec{n}$ is the direction normal to DW) remain almost motionless. In contrast, the back and front DWs perpendicular to the current (i.e., $\vec{j} \parallel \vec{n}$) are significantly displaced. Surprisingly, the back DW moves faster than the front DW, which causes the collapse of the domain (see Fig. ~\ref{fig:5} (E-F)). 
Another interesting feature is the increasingly pointed shape of the front DW (not observed for the back DW). Here, the faceting of the front DW develops without any contribution of "strong" pinning sites \LPSv{(see~\cite{Moon_PRL_2013,supplemental}),} which suggests that the transverse orientation between DW and current is unstable. Consequently, the different shape evolutions of the front and back DWs observed in Fig. ~\ref{fig:5} (A-F) can be interpreted as a result of opposite contributions of DW elasticity. The two side DWs pull the extremities of back (front) DWs in the  $-\vec{j}$ ($\vec{j}$) direction, which tends to stabilize (destabilize) the transverse DW orientation. 

\LPSv{To explore more quantitatively the directionality of interaction between DW and current, we have measured the DW displacements $\Delta x_n$ along the direction $\vec{n}$ as a function of the angle $\theta$ between  $\vec{n}$ and $-\vec{j}$,
for a fixed magnitude of current density and pulse duration $\Delta t=1\mu s$ (see Fig. ~\ref{fig:5} (G and H)). We have then deduced the variation of velocity $v_n=\Delta x_n/\Delta t$ as a function of $\left|j\right| \cos \theta$, which are reported Fig. ~\ref{fig:5} (I).
As expected the DW velocity decreases as the tilting angle increases (i.e.,$\left|j\right| cos \theta$ decreases). For $\left|j\right| \cos \theta<2.25~GA/m^2$, the displacement $\Delta x_n$ becomes lower than the spatial resolution 
so that the estimation for $v_n$ is zero. The dispersion of data most probably results from the contributions of DW elasticity, which tends to reduce DW velocity. Therefore, the velocity DW without contribution of elasticity should correspond to the upper measured values. Interestingly, those values present a rather good overlap with the velocity curves of Fig.\ref{fig:3}, which were obtained as a function of the current density and for $\theta =0$. 
This indicates that DW tilting reduces the driving force in the direction normal $\vec{n}$ to the DW and suggests that the DW faceting originates from the directionality of driving force $f \propto  \vec{j} \cdot \vec{n}$.}

\LPSv{Let us now discuss the implication of the variation of the driving force with DW tilting ($\propto j\cos~\theta$) on universal behaviors.} 
The second order expansion of $cos~\theta$ ($\approx 1-(1/2)(\partial u/\partial x)^2$) introduces a so-called KPZ term ($\lambda (\partial u/\partial x)^2$ with $\lambda=-j/2$) in the equation of motion~\cite{kardar_PRL_1986_dynamic}, which is usually used to describe interface tilting. However, at large tilting angle, the relation $f \propto  \vec{j} \cdot \vec{n}$ should not be compatible with the KPZ minimal model.   
In contrast, in the direction $\vec{n}$ perpendicular to DW, for a fixed DW tilting, the driving force due to spin-transfer-torque is simply proportional to $ j\cos~\theta$ and acts as a magnetic field ($f \propto H$). This could explain the compatibility of current induced DW motion with the qEW universality class, which is observed experimentally.

In conclusion, the common universal behaviors with field driven DW motion and in particular the agreement with the self-consistent description of creep and depinning regimes could help to better understand spin-transfer-torque induced DW motion experiments~\cite{caretta_natnano_2018} since it allows a clear identification of dynamical regimes and to discriminate universal from material dependent behaviors~\cite{jeudy_PRB_2018_DW_pinning}. It would be also very interesting to study the universal behaviors of DWs driven by spin-orbit-torque~\cite{martinez_APL_2013} and of DWs in antiferromagnets~\cite{yang_naturenano_2015}.
Moreover, the instability of transverse alignment between DW and current should have direct implications for potential applications based on the controlled motion of DWs~\cite{parkin_naturenano_2015_race_track} 
in nanowires.
%

\begin{acknowledgments}
We wish to thank S. Bustingorry, A. Kolton, and K. Wiese for fruitful discussions. This work was partly supported by the projects DIM CNano IdF (Region Ile-de-France) and the Labex NanoSaclay (ANR-10-LABX-0035). R.D.P. thanks the Mexican council CONACyT for the PhD fellowship n0: 449563.
\end{acknowledgments}

\bibliography{refs_para_I,refs_para_old_2018,comments}

\begin{thebibliography}{33}%
\makeatletter
\providecommand \@ifxundefined [1]{%
 \@ifx{#1\undefined}
}%
\providecommand \@ifnum [1]{%
 \ifnum #1\expandafter \@firstoftwo
 \else \expandafter \@secondoftwo
 \fi
}%
\providecommand \@ifx [1]{%
 \ifx #1\expandafter \@firstoftwo
 \else \expandafter \@secondoftwo
 \fi
}%
\providecommand \natexlab [1]{#1}%
\providecommand \enquote  [1]{``#1''}%
\providecommand \bibnamefont  [1]{#1}%
\providecommand \bibfnamefont [1]{#1}%
\providecommand \citenamefont [1]{#1}%
\providecommand \href@noop [0]{\@secondoftwo}%
\providecommand \href [0]{\begingroup \@sanitize@url \@href}%
\providecommand \@href[1]{\@@startlink{#1}\@@href}%
\providecommand \@@href[1]{\endgroup#1\@@endlink}%
\providecommand \@sanitize@url [0]{\catcode `\\12\catcode `\$12\catcode
  `\&12\catcode `\#12\catcode `\^12\catcode `\_12\catcode `\%12\relax}%
\providecommand \@@startlink[1]{}%
\providecommand \@@endlink[0]{}%
\providecommand \url  [0]{\begingroup\@sanitize@url \@url }%
\providecommand \@url [1]{\endgroup\@href {#1}{\urlprefix }}%
\providecommand \urlprefix  [0]{URL }%
\providecommand \Eprint [0]{\href }%
\providecommand \doibase [0]{http://dx.doi.org/}%
\providecommand \selectlanguage [0]{\@gobble}%
\providecommand \bibinfo  [0]{\@secondoftwo}%
\providecommand \bibfield  [0]{\@secondoftwo}%
\providecommand \translation [1]{[#1]}%
\providecommand \BibitemOpen [0]{}%
\providecommand \bibitemStop [0]{}%
\providecommand \bibitemNoStop [0]{.\EOS\space}%
\providecommand \EOS [0]{\spacefactor3000\relax}%
\providecommand \BibitemShut  [1]{\csname bibitem#1\endcsname}%
\let\auto@bib@innerbib\@empty
\bibitem [{\citenamefont {Parkin}\ and\ \citenamefont
  {Yang}(2015)}]{parkin_naturenano_2015_race_track}%
  \BibitemOpen
  \bibfield  {author} {\bibinfo {author} {\bibfnamefont {S.~S.~P.}\
  \bibnamefont {Parkin}}\ and\ \bibinfo {author} {\bibfnamefont {S.-H.}\
  \bibnamefont {Yang}},\ }\href {\doibase 10.1038/nnano.2015.41} {\bibfield
  {journal} {\bibinfo  {journal} {Nature Nanotechnology}\ }\textbf {\bibinfo
  {volume} {10}},\ \bibinfo {pages} {195} (\bibinfo {year} {2015})}\BibitemShut
  {NoStop}%
\bibitem [{\citenamefont {Caretta}\ \emph {et~al.}(2018)\citenamefont
  {Caretta}, \citenamefont {Mann}, \citenamefont {Büttner}, \citenamefont
  {Ueda}, \citenamefont {Pfau}, \citenamefont {Günther}, \citenamefont
  {Hessing}, \citenamefont {Churikova}, \citenamefont {Klose}, \citenamefont
  {Schneider}, \citenamefont {Engel}, \citenamefont {Marcus}, \citenamefont
  {Bono}, \citenamefont {Bagschik}, \citenamefont {Eisebitt},\ and\
  \citenamefont {Beach}}]{caretta_natnano_2018}%
  \BibitemOpen
  \bibfield  {author} {\bibinfo {author} {\bibfnamefont {L.}~\bibnamefont
  {Caretta}}, \bibinfo {author} {\bibfnamefont {M.}~\bibnamefont {Mann}},
  \bibinfo {author} {\bibfnamefont {F.}~\bibnamefont {Büttner}}, \bibinfo
  {author} {\bibfnamefont {K.}~\bibnamefont {Ueda}}, \bibinfo {author}
  {\bibfnamefont {B.}~\bibnamefont {Pfau}}, \bibinfo {author} {\bibfnamefont
  {C.~M.}\ \bibnamefont {Günther}}, \bibinfo {author} {\bibfnamefont
  {P.}~\bibnamefont {Hessing}}, \bibinfo {author} {\bibfnamefont
  {A.}~\bibnamefont {Churikova}}, \bibinfo {author} {\bibfnamefont
  {C.}~\bibnamefont {Klose}}, \bibinfo {author} {\bibfnamefont
  {M.}~\bibnamefont {Schneider}}, \bibinfo {author} {\bibfnamefont
  {D.}~\bibnamefont {Engel}}, \bibinfo {author} {\bibfnamefont
  {C.}~\bibnamefont {Marcus}}, \bibinfo {author} {\bibfnamefont
  {D.}~\bibnamefont {Bono}}, \bibinfo {author} {\bibfnamefont {K.}~\bibnamefont
  {Bagschik}}, \bibinfo {author} {\bibfnamefont {S.}~\bibnamefont {Eisebitt}},
  \ and\ \bibinfo {author} {\bibfnamefont {G.~S.~D.}\ \bibnamefont {Beach}},\
  }\href@noop {} {\bibfield  {journal} {\bibinfo  {journal} {Nature
  Nanotechnology}\ ,\ \bibinfo {pages} {1}} (\bibinfo {year}
  {2018})}\BibitemShut {NoStop}%
\bibitem [{\citenamefont {Hrabec}\ \emph {et~al.}(2019)\citenamefont {Hrabec},
  \citenamefont {Shahbazi}, \citenamefont {Moore}, \citenamefont {Martinez},\
  and\ \citenamefont {Marrows}}]{hrabec_nanotech_2019}%
  \BibitemOpen
  \bibfield  {author} {\bibinfo {author} {\bibfnamefont {A.}~\bibnamefont
  {Hrabec}}, \bibinfo {author} {\bibfnamefont {K.}~\bibnamefont {Shahbazi}},
  \bibinfo {author} {\bibfnamefont {T.~A.}\ \bibnamefont {Moore}}, \bibinfo
  {author} {\bibfnamefont {E.}~\bibnamefont {Martinez}}, \ and\ \bibinfo
  {author} {\bibfnamefont {C.~H.}\ \bibnamefont {Marrows}},\ }\href {\doibase
  10.1088/1361-6528/ab087b} {\bibfield  {journal} {\bibinfo  {journal}
  {Nanotechnology}\ }\textbf {\bibinfo {volume} {30}},\ \bibinfo {pages}
  {234003} (\bibinfo {year} {2019})}\BibitemShut {NoStop}%
\bibitem [{\citenamefont {Lemerle}\ \emph {et~al.}(1998)\citenamefont
  {Lemerle}, \citenamefont {Ferr{\'e}}, \citenamefont {Chappert}, \citenamefont
  {Mathet}, \citenamefont {Giamarchi},\ and\ \citenamefont {{Le
  Doussal}}}]{lemerle_PRL_1998_domainwall_creep}%
  \BibitemOpen
  \bibfield  {author} {\bibinfo {author} {\bibfnamefont {S.}~\bibnamefont
  {Lemerle}}, \bibinfo {author} {\bibfnamefont {J.}~\bibnamefont {Ferr{\'e}}},
  \bibinfo {author} {\bibfnamefont {C.}~\bibnamefont {Chappert}}, \bibinfo
  {author} {\bibfnamefont {V.}~\bibnamefont {Mathet}}, \bibinfo {author}
  {\bibfnamefont {T.}~\bibnamefont {Giamarchi}}, \ and\ \bibinfo {author}
  {\bibfnamefont {P.}~\bibnamefont {{Le Doussal}}},\ }\href@noop {} {\bibfield
  {journal} {\bibinfo  {journal} {Phys. Rev. Lett.}\ }\textbf {\bibinfo
  {volume} {80}},\ \bibinfo {pages} {849} (\bibinfo {year} {1998})}\BibitemShut
  {NoStop}%
\bibitem [{\citenamefont {Moon}\ \emph {et~al.}(2013)\citenamefont {Moon},
  \citenamefont {Kim}, \citenamefont {Yoo}, \citenamefont {Cho}, \citenamefont
  {Hwang}, \citenamefont {Kahng}, \citenamefont {Min}, \citenamefont {Shin},\
  and\ \citenamefont {Choe}}]{Moon_PRL_2013}%
  \BibitemOpen
  \bibfield  {author} {\bibinfo {author} {\bibfnamefont {K.-W.}\ \bibnamefont
  {Moon}}, \bibinfo {author} {\bibfnamefont {D.-H.}\ \bibnamefont {Kim}},
  \bibinfo {author} {\bibfnamefont {S.-C.}\ \bibnamefont {Yoo}}, \bibinfo
  {author} {\bibfnamefont {C.-G.}\ \bibnamefont {Cho}}, \bibinfo {author}
  {\bibfnamefont {S.}~\bibnamefont {Hwang}}, \bibinfo {author} {\bibfnamefont
  {B.}~\bibnamefont {Kahng}}, \bibinfo {author} {\bibfnamefont {B.-C.}\
  \bibnamefont {Min}}, \bibinfo {author} {\bibfnamefont {K.-H.}\ \bibnamefont
  {Shin}}, \ and\ \bibinfo {author} {\bibfnamefont {S.-B.}\ \bibnamefont
  {Choe}},\ }\href {\doibase doi.org/10.1103/PhysRevLett.110.139902} {\bibfield
   {journal} {\bibinfo  {journal} {Phys. Rev. Lett.}\ }\textbf {\bibinfo
  {volume} {110}},\ \bibinfo {pages} {107203} (\bibinfo {year}
  {2013})}\BibitemShut {NoStop}%
\bibitem [{\citenamefont {Ferrero}\ \emph {et~al.}(2017)\citenamefont
  {Ferrero}, \citenamefont {Foini}, \citenamefont {Giamarchi}, \citenamefont
  {Kolton},\ and\ \citenamefont
  {Rosso}}]{ferrero_prl_2017_spatiotemporal_patterns}%
  \BibitemOpen
  \bibfield  {author} {\bibinfo {author} {\bibfnamefont {E.~E.}\ \bibnamefont
  {Ferrero}}, \bibinfo {author} {\bibfnamefont {L.}~\bibnamefont {Foini}},
  \bibinfo {author} {\bibfnamefont {T.}~\bibnamefont {Giamarchi}}, \bibinfo
  {author} {\bibfnamefont {A.~B.}\ \bibnamefont {Kolton}}, \ and\ \bibinfo
  {author} {\bibfnamefont {A.}~\bibnamefont {Rosso}},\ }\href {\doibase
  10.1103/PhysRevLett.118.147208} {\bibfield  {journal} {\bibinfo  {journal}
  {Phys. Rev. Lett.}\ }\textbf {\bibinfo {volume} {118}},\ \bibinfo {pages}
  {147208} (\bibinfo {year} {2017})}\BibitemShut {NoStop}%
\bibitem [{\citenamefont {Grassi}\ \emph {et~al.}(2018)\citenamefont {Grassi},
  \citenamefont {Kolton}, \citenamefont {Jeudy}, \citenamefont {Mougin},
  \citenamefont {Bustingorry},\ and\ \citenamefont
  {Curiale}}]{grassi_prb_2018}%
  \BibitemOpen
  \bibfield  {author} {\bibinfo {author} {\bibfnamefont {M.~P.}\ \bibnamefont
  {Grassi}}, \bibinfo {author} {\bibfnamefont {A.~B.}\ \bibnamefont {Kolton}},
  \bibinfo {author} {\bibfnamefont {V.}~\bibnamefont {Jeudy}}, \bibinfo
  {author} {\bibfnamefont {A.}~\bibnamefont {Mougin}}, \bibinfo {author}
  {\bibfnamefont {S.}~\bibnamefont {Bustingorry}}, \ and\ \bibinfo {author}
  {\bibfnamefont {J.}~\bibnamefont {Curiale}},\ }\href {\doibase
  10.1103/PhysRevB.98.224201} {\bibfield  {journal} {\bibinfo  {journal} {Phys.
  Rev. B}\ }\textbf {\bibinfo {volume} {98}},\ \bibinfo {pages} {224201}
  (\bibinfo {year} {2018})}\BibitemShut {NoStop}%
\bibitem [{\citenamefont {Yamanouchi}\ \emph {et~al.}(2007)\citenamefont
  {Yamanouchi}, \citenamefont {Ieda}, \citenamefont {Matsukura}, \citenamefont
  {Barnes}, \citenamefont {Maekawa},\ and\ \citenamefont
  {Ohno}}]{Yamanouchi_science_2007}%
  \BibitemOpen
  \bibfield  {author} {\bibinfo {author} {\bibfnamefont {M.}~\bibnamefont
  {Yamanouchi}}, \bibinfo {author} {\bibfnamefont {J.}~\bibnamefont {Ieda}},
  \bibinfo {author} {\bibfnamefont {F.}~\bibnamefont {Matsukura}}, \bibinfo
  {author} {\bibfnamefont {S.~E.}\ \bibnamefont {Barnes}}, \bibinfo {author}
  {\bibfnamefont {S.}~\bibnamefont {Maekawa}}, \ and\ \bibinfo {author}
  {\bibfnamefont {H.}~\bibnamefont {Ohno}},\ }\href {\doibase
  10.1126/science.1145516} {\bibfield  {journal} {\bibinfo  {journal}
  {Science}\ }\textbf {\bibinfo {volume} {317}},\ \bibinfo {pages} {1726}
  (\bibinfo {year} {2007})}\BibitemShut {NoStop}%
\bibitem [{\citenamefont {Metaxas}\ \emph {et~al.}(2007)\citenamefont
  {Metaxas}, \citenamefont {Jamet}, \citenamefont {Mougin}, \citenamefont
  {Cormier}, \citenamefont {Ferr{\'e}}, \citenamefont {Baltz}, \citenamefont
  {Rodmacq}, \citenamefont {Dieny},\ and\ \citenamefont
  {Stamps}}]{metaxas_PRL_07_depinning_thermal_rounding}%
  \BibitemOpen
  \bibfield  {author} {\bibinfo {author} {\bibfnamefont {P.~J.}\ \bibnamefont
  {Metaxas}}, \bibinfo {author} {\bibfnamefont {J.~P.}\ \bibnamefont {Jamet}},
  \bibinfo {author} {\bibfnamefont {A.}~\bibnamefont {Mougin}}, \bibinfo
  {author} {\bibfnamefont {M.}~\bibnamefont {Cormier}}, \bibinfo {author}
  {\bibfnamefont {J.}~\bibnamefont {Ferr{\'e}}}, \bibinfo {author}
  {\bibfnamefont {V.}~\bibnamefont {Baltz}}, \bibinfo {author} {\bibfnamefont
  {B.}~\bibnamefont {Rodmacq}}, \bibinfo {author} {\bibfnamefont
  {B.}~\bibnamefont {Dieny}}, \ and\ \bibinfo {author} {\bibfnamefont {R.~L.}\
  \bibnamefont {Stamps}},\ }\href@noop {} {\bibfield  {journal} {\bibinfo
  {journal} {Phys. Rev. Lett.}\ }\textbf {\bibinfo {volume} {99}},\ \bibinfo
  {pages} {217208} (\bibinfo {year} {2007})}\BibitemShut {NoStop}%
\bibitem [{\citenamefont {DuttaGupta}\ \emph {et~al.}(2016)\citenamefont
  {DuttaGupta}, \citenamefont {Fukami}, \citenamefont {Zhang}, \citenamefont
  {Sato}, \citenamefont {Yamanouchi}, \citenamefont {Matsukura},\ and\
  \citenamefont {Ohno}}]{duttagupta_natphys_2015}%
  \BibitemOpen
  \bibfield  {author} {\bibinfo {author} {\bibfnamefont {S.}~\bibnamefont
  {DuttaGupta}}, \bibinfo {author} {\bibfnamefont {S.}~\bibnamefont {Fukami}},
  \bibinfo {author} {\bibfnamefont {C.}~\bibnamefont {Zhang}}, \bibinfo
  {author} {\bibfnamefont {H.}~\bibnamefont {Sato}}, \bibinfo {author}
  {\bibfnamefont {M.}~\bibnamefont {Yamanouchi}}, \bibinfo {author}
  {\bibfnamefont {F.}~\bibnamefont {Matsukura}}, \ and\ \bibinfo {author}
  {\bibfnamefont {H.}~\bibnamefont {Ohno}},\ }\href@noop {} {\bibfield
  {journal} {\bibinfo  {journal} {Nature Physics}\ }\textbf {\bibinfo {volume}
  {12}},\ \bibinfo {pages} {333} (\bibinfo {year} {2016})}\BibitemShut
  {NoStop}%
\bibitem [{\citenamefont {Jeudy}\ \emph {et~al.}(2016)\citenamefont {Jeudy},
  \citenamefont {Mougin}, \citenamefont {Bustingorry}, \citenamefont
  {Savero~Torres}, \citenamefont {Gorchon}, \citenamefont {Kolton},
  \citenamefont {Lema\^{\i}tre},\ and\ \citenamefont
  {Jamet}}]{jeudy_PRL_2016_energy_barrier}%
  \BibitemOpen
  \bibfield  {author} {\bibinfo {author} {\bibfnamefont {V.}~\bibnamefont
  {Jeudy}}, \bibinfo {author} {\bibfnamefont {A.}~\bibnamefont {Mougin}},
  \bibinfo {author} {\bibfnamefont {S.}~\bibnamefont {Bustingorry}}, \bibinfo
  {author} {\bibfnamefont {W.}~\bibnamefont {Savero~Torres}}, \bibinfo {author}
  {\bibfnamefont {J.}~\bibnamefont {Gorchon}}, \bibinfo {author} {\bibfnamefont
  {A.~B.}\ \bibnamefont {Kolton}}, \bibinfo {author} {\bibfnamefont
  {A.}~\bibnamefont {Lema\^{\i}tre}}, \ and\ \bibinfo {author} {\bibfnamefont
  {J.-P.}\ \bibnamefont {Jamet}},\ }\href {\doibase
  10.1103/PhysRevLett.117.057201} {\bibfield  {journal} {\bibinfo  {journal}
  {Phys. Rev. Lett.}\ }\textbf {\bibinfo {volume} {117}},\ \bibinfo {pages}
  {057201} (\bibinfo {year} {2016})}\BibitemShut {NoStop}%
\bibitem [{\citenamefont {Diaz~Pardo}\ \emph {et~al.}(2017)\citenamefont
  {Diaz~Pardo}, \citenamefont {Savero~Torres}, \citenamefont {Kolton},
  \citenamefont {Bustingorry},\ and\ \citenamefont
  {Jeudy}}]{diaz_PRB_2017_depinning}%
  \BibitemOpen
  \bibfield  {author} {\bibinfo {author} {\bibfnamefont {R.}~\bibnamefont
  {Diaz~Pardo}}, \bibinfo {author} {\bibfnamefont {W.}~\bibnamefont
  {Savero~Torres}}, \bibinfo {author} {\bibfnamefont {A.~B.}\ \bibnamefont
  {Kolton}}, \bibinfo {author} {\bibfnamefont {S.}~\bibnamefont {Bustingorry}},
  \ and\ \bibinfo {author} {\bibfnamefont {V.}~\bibnamefont {Jeudy}},\ }\href
  {\doibase 10.1103/PhysRevB.95.184434} {\bibfield  {journal} {\bibinfo
  {journal} {Phys. Rev. B}\ }\textbf {\bibinfo {volume} {95}},\ \bibinfo
  {pages} {184434} (\bibinfo {year} {2017})}\BibitemShut {NoStop}%
\bibitem [{\citenamefont {Atis}\ \emph {et~al.}(2015)\citenamefont {Atis},
  \citenamefont {Dubey}, \citenamefont {Salin}, \citenamefont {Talon},
  \citenamefont {Le~Doussal},\ and\ \citenamefont {Wiese}}]{atis_prl_2015}%
  \BibitemOpen
  \bibfield  {author} {\bibinfo {author} {\bibfnamefont {S.}~\bibnamefont
  {Atis}}, \bibinfo {author} {\bibfnamefont {A.~K.}\ \bibnamefont {Dubey}},
  \bibinfo {author} {\bibfnamefont {D.}~\bibnamefont {Salin}}, \bibinfo
  {author} {\bibfnamefont {L.}~\bibnamefont {Talon}}, \bibinfo {author}
  {\bibfnamefont {P.}~\bibnamefont {Le~Doussal}}, \ and\ \bibinfo {author}
  {\bibfnamefont {K.~J.}\ \bibnamefont {Wiese}},\ }\href {\doibase
  10.1103/PhysRevLett.114.234502} {\bibfield  {journal} {\bibinfo  {journal}
  {Phys. Rev. Lett.}\ }\textbf {\bibinfo {volume} {114}},\ \bibinfo {pages}
  {234502} (\bibinfo {year} {2015})}\BibitemShut {NoStop}%
\bibitem [{\citenamefont {Bonachela}\ \emph {et~al.}(2011)\citenamefont
  {Bonachela}, \citenamefont {Nadell}, \citenamefont {Xavier},\ and\
  \citenamefont {Levin}}]{Bonachela_bacteria_colonies_2011}%
  \BibitemOpen
  \bibfield  {author} {\bibinfo {author} {\bibfnamefont {J.~A.}\ \bibnamefont
  {Bonachela}}, \bibinfo {author} {\bibfnamefont {C.~D.}\ \bibnamefont
  {Nadell}}, \bibinfo {author} {\bibfnamefont {J.~B.}\ \bibnamefont {Xavier}},
  \ and\ \bibinfo {author} {\bibfnamefont {S.~A.}\ \bibnamefont {Levin}},\
  }\href {\doibase 10.1007/s10955-011-0179-x} {\bibfield  {journal} {\bibinfo
  {journal} {Journal of Statistical Physics}\ }\textbf {\bibinfo {volume}
  {144}},\ \bibinfo {pages} {303} (\bibinfo {year} {2011})}\BibitemShut
  {NoStop}%
\bibitem [{\citenamefont {Moulinet}\ \emph {et~al.}(2002)\citenamefont
  {Moulinet}, \citenamefont {Guthmann},\ and\ \citenamefont
  {Rolley}}]{moulinet_EPJE_2002}%
  \BibitemOpen
  \bibfield  {author} {\bibinfo {author} {\bibfnamefont {S.}~\bibnamefont
  {Moulinet}}, \bibinfo {author} {\bibfnamefont {C.}~\bibnamefont {Guthmann}},
  \ and\ \bibinfo {author} {\bibfnamefont {E.}~\bibnamefont {Rolley}},\ }\href
  {\doibase 10.1140/epje/i2002-10032-2} {\bibfield  {journal} {\bibinfo
  {journal} {The European Physical Journal E}\ }\textbf {\bibinfo {volume}
  {8}},\ \bibinfo {pages} {437} (\bibinfo {year} {2002})}\BibitemShut {NoStop}%
\bibitem [{\citenamefont {Tybell}\ \emph {et~al.}(2002)\citenamefont {Tybell},
  \citenamefont {Paruch}, \citenamefont {Giamarchi},\ and\ \citenamefont
  {Triscone}}]{tybell2002}%
  \BibitemOpen
  \bibfield  {author} {\bibinfo {author} {\bibfnamefont {T.}~\bibnamefont
  {Tybell}}, \bibinfo {author} {\bibfnamefont {P.}~\bibnamefont {Paruch}},
  \bibinfo {author} {\bibfnamefont {T.}~\bibnamefont {Giamarchi}}, \ and\
  \bibinfo {author} {\bibfnamefont {J.-M.}\ \bibnamefont {Triscone}},\ }\href
  {\doibase 10.1103/PhysRevLett.89.097601} {\bibfield  {journal} {\bibinfo
  {journal} {Phys. Rev. Lett.}\ }\textbf {\bibinfo {volume} {89}},\ \bibinfo
  {pages} {097601} (\bibinfo {year} {2002})}\BibitemShut {NoStop}%
\bibitem [{\citenamefont {Chauve}\ \emph {et~al.}(2000)\citenamefont {Chauve},
  \citenamefont {Giamarchi},\ and\ \citenamefont {Le~Doussal}}]{chauve_2000}%
  \BibitemOpen
  \bibfield  {author} {\bibinfo {author} {\bibfnamefont {P.}~\bibnamefont
  {Chauve}}, \bibinfo {author} {\bibfnamefont {T.}~\bibnamefont {Giamarchi}}, \
  and\ \bibinfo {author} {\bibfnamefont {P.}~\bibnamefont {Le~Doussal}},\
  }\href {\doibase 10.1103/PhysRevB.62.6241} {\bibfield  {journal} {\bibinfo
  {journal} {Phys. Rev. B}\ }\textbf {\bibinfo {volume} {62}},\ \bibinfo
  {pages} {6241} (\bibinfo {year} {2000})}\BibitemShut {NoStop}%
\bibitem [{\citenamefont {{Le Doussal}}\ and\ \citenamefont
  {Wiese}(2009)}]{LeDoussal2009}%
  \BibitemOpen
  \bibfield  {author} {\bibinfo {author} {\bibfnamefont {P.}~\bibnamefont {{Le
  Doussal}}}\ and\ \bibinfo {author} {\bibfnamefont {K.~J.}\ \bibnamefont
  {Wiese}},\ }\href@noop {} {\bibfield  {journal} {\bibinfo  {journal}
  {Physical Review E - Statistical, Nonlinear, and Soft Matter Physics}\
  }\textbf {\bibinfo {volume} {79}} (\bibinfo {year} {2009})},\ \Eprint
  {http://arxiv.org/abs/0808.3217} {0808.3217} \BibitemShut {NoStop}%
\bibitem [{\citenamefont {Bustingorry}\ \emph {et~al.}(2012)\citenamefont
  {Bustingorry}, \citenamefont {Kolton},\ and\ \citenamefont
  {Giamarchi}}]{bustingorry_PRE_12_thermal_rounding}%
  \BibitemOpen
  \bibfield  {author} {\bibinfo {author} {\bibfnamefont {S.}~\bibnamefont
  {Bustingorry}}, \bibinfo {author} {\bibfnamefont {A.~B.}\ \bibnamefont
  {Kolton}}, \ and\ \bibinfo {author} {\bibfnamefont {T.}~\bibnamefont
  {Giamarchi}},\ }\href {\doibase 10.1103/PhysRevE.85.021144} {\bibfield
  {journal} {\bibinfo  {journal} {Phys. Rev. E}\ }\textbf {\bibinfo {volume}
  {85}},\ \bibinfo {pages} {021144} (\bibinfo {year} {2012})}\BibitemShut
  {NoStop}%
\bibitem [{\citenamefont {Rosso}\ and\ \citenamefont
  {Krauth}(2001)}]{rosso_prl_2001}%
  \BibitemOpen
  \bibfield  {author} {\bibinfo {author} {\bibfnamefont {A.}~\bibnamefont
  {Rosso}}\ and\ \bibinfo {author} {\bibfnamefont {W.}~\bibnamefont {Krauth}},\
  }\href {\doibase 10.1103/PhysRevLett.87.187002} {\bibfield  {journal}
  {\bibinfo  {journal} {Phys. Rev. Lett.}\ }\textbf {\bibinfo {volume} {87}},\
  \bibinfo {pages} {187002} (\bibinfo {year} {2001})}\BibitemShut {NoStop}%
\bibitem [{\citenamefont {Kolton}\ \emph {et~al.}(2009)\citenamefont {Kolton},
  \citenamefont {Rosso}, \citenamefont {Giamarchi},\ and\ \citenamefont
  {Krauth}}]{kolton_prb_2009_pathways}%
  \BibitemOpen
  \bibfield  {author} {\bibinfo {author} {\bibfnamefont {A.~B.}\ \bibnamefont
  {Kolton}}, \bibinfo {author} {\bibfnamefont {A.}~\bibnamefont {Rosso}},
  \bibinfo {author} {\bibfnamefont {T.}~\bibnamefont {Giamarchi}}, \ and\
  \bibinfo {author} {\bibfnamefont {W.}~\bibnamefont {Krauth}},\ }\href@noop {}
  {\bibfield  {journal} {\bibinfo  {journal} {Phys. Rev. B}\ }\textbf {\bibinfo
  {volume} {79}},\ \bibinfo {pages} {184207} (\bibinfo {year}
  {2009})}\BibitemShut {NoStop}%
\bibitem [{\citenamefont {Lee}\ \emph {et~al.}(2011)\citenamefont {Lee},
  \citenamefont {Kim}, \citenamefont {Ryu}, \citenamefont {Moon}, \citenamefont
  {Yun}, \citenamefont {Gim}, \citenamefont {Lee}, \citenamefont {Shin},
  \citenamefont {Lee},\ and\ \citenamefont {Choe}}]{lee_prl_2011}%
  \BibitemOpen
  \bibfield  {author} {\bibinfo {author} {\bibfnamefont {J.-C.}\ \bibnamefont
  {Lee}}, \bibinfo {author} {\bibfnamefont {K.-J.}\ \bibnamefont {Kim}},
  \bibinfo {author} {\bibfnamefont {J.}~\bibnamefont {Ryu}}, \bibinfo {author}
  {\bibfnamefont {K.-W.}\ \bibnamefont {Moon}}, \bibinfo {author}
  {\bibfnamefont {S.-J.}\ \bibnamefont {Yun}}, \bibinfo {author} {\bibfnamefont
  {G.-H.}\ \bibnamefont {Gim}}, \bibinfo {author} {\bibfnamefont {K.-S.}\
  \bibnamefont {Lee}}, \bibinfo {author} {\bibfnamefont {K.-H.}\ \bibnamefont
  {Shin}}, \bibinfo {author} {\bibfnamefont {H.-W.}\ \bibnamefont {Lee}}, \
  and\ \bibinfo {author} {\bibfnamefont {S.-B.}\ \bibnamefont {Choe}},\ }\href
  {\doibase 10.1103/PhysRevLett.107.067201} {\bibfield  {journal} {\bibinfo
  {journal} {Phys. Rev. Lett.}\ }\textbf {\bibinfo {volume} {107}},\ \bibinfo
  {pages} {067201} (\bibinfo {year} {2011})}\BibitemShut {NoStop}%
\bibitem [{\citenamefont {Moon}\ \emph {et~al.}(2018)\citenamefont {Moon},
  \citenamefont {Kim}, \citenamefont {Yoon}, \citenamefont {Choi},
  \citenamefont {Kim}, \citenamefont {Song}, \citenamefont {Kim}, \citenamefont
  {Chun},\ and\ \citenamefont {Hwang}}]{Moon_natcomm_2018}%
  \BibitemOpen
  \bibfield  {author} {\bibinfo {author} {\bibfnamefont {K.-W.}\ \bibnamefont
  {Moon}}, \bibinfo {author} {\bibfnamefont {C.}~\bibnamefont {Kim}}, \bibinfo
  {author} {\bibfnamefont {J.}~\bibnamefont {Yoon}}, \bibinfo {author}
  {\bibfnamefont {J.~W.}\ \bibnamefont {Choi}}, \bibinfo {author}
  {\bibfnamefont {D.-O.}\ \bibnamefont {Kim}}, \bibinfo {author} {\bibfnamefont
  {K.~M.}\ \bibnamefont {Song}}, \bibinfo {author} {\bibfnamefont
  {D.}~\bibnamefont {Kim}}, \bibinfo {author} {\bibfnamefont {B.~S.}\
  \bibnamefont {Chun}}, \ and\ \bibinfo {author} {\bibfnamefont
  {C.}~\bibnamefont {Hwang}},\ }\href {\doibase 10.1038/s41467-018-06223-z}
  {\bibfield  {journal} {\bibinfo  {journal} {Nature Communications}\ }\textbf
  {\bibinfo {volume} {9}},\ \bibinfo {pages} {3788} (\bibinfo {year}
  {2018})}\BibitemShut {NoStop}%
\bibitem [{\citenamefont {Niazi}\ \emph {et~al.}(2013)\citenamefont {Niazi},
  \citenamefont {Cormier}, \citenamefont {Lucot}, \citenamefont {Largeau},
  \citenamefont {Jeudy}, \citenamefont {Cibert},\ and\ \citenamefont
  {Lema\^{i}tre}}]{niazi_apl_2013}%
  \BibitemOpen
  \bibfield  {author} {\bibinfo {author} {\bibfnamefont {T.}~\bibnamefont
  {Niazi}}, \bibinfo {author} {\bibfnamefont {M.}~\bibnamefont {Cormier}},
  \bibinfo {author} {\bibfnamefont {D.}~\bibnamefont {Lucot}}, \bibinfo
  {author} {\bibfnamefont {L.}~\bibnamefont {Largeau}}, \bibinfo {author}
  {\bibfnamefont {V.}~\bibnamefont {Jeudy}}, \bibinfo {author} {\bibfnamefont
  {J.}~\bibnamefont {Cibert}}, \ and\ \bibinfo {author} {\bibfnamefont
  {A.}~\bibnamefont {Lema\^{i}tre}},\ }\href {\doibase 10.1063/1.4798258}
  {\bibfield  {journal} {\bibinfo  {journal} {Applied Physics Letters}\
  }\textbf {\bibinfo {volume} {102}},\ \bibinfo {pages} {122403} (\bibinfo
  {year} {2013})}\BibitemShut {NoStop}%
\bibitem [{sup()}]{supplemental}%
  \BibitemOpen
  \href@noop {} {\emph {\bibinfo {title} {See Supplemental Material at
  http://link.aps.org/ supplemental/... for details on the sample, Joule
  heating studies, measurements of velocity and roughness
  exponent..}}}\BibitemShut {Stop}%
\bibitem [{com({\natexlab{a}})}]{comment}%
  \BibitemOpen
  \href@noop {} {\emph {\bibinfo {title} {This value of $\mu_j$ is compatible
  with measurements performed using Pt/Co/Pt nanowires~\cite{lee_prl_2011}. The
  largest values reported for $\mu_j$ and $\mu_H$ in
  Ref.~\onlinecite{Yamanouchi_science_2007} for (Ga,Mn)As could not be
  reproduced. See also Ref. \cite{jeudy_PRL_2016_energy_barrier} for a
  discussion.}}}\BibitemShut {Stop}%
\bibitem [{\citenamefont {Jeudy}\ \emph {et~al.}(2018)\citenamefont {Jeudy},
  \citenamefont {D\'{\i}az~Pardo}, \citenamefont {Savero~Torres}, \citenamefont
  {Bustingorry},\ and\ \citenamefont {Kolton}}]{jeudy_PRB_2018_DW_pinning}%
  \BibitemOpen
  \bibfield  {author} {\bibinfo {author} {\bibfnamefont {V.}~\bibnamefont
  {Jeudy}}, \bibinfo {author} {\bibfnamefont {R.}~\bibnamefont
  {D\'{\i}az~Pardo}}, \bibinfo {author} {\bibfnamefont {W.}~\bibnamefont
  {Savero~Torres}}, \bibinfo {author} {\bibfnamefont {S.}~\bibnamefont
  {Bustingorry}}, \ and\ \bibinfo {author} {\bibfnamefont {A.~B.}\ \bibnamefont
  {Kolton}},\ }\href {\doibase 10.1103/PhysRevB.98.054406} {\bibfield
  {journal} {\bibinfo  {journal} {Phys. Rev. B}\ }\textbf {\bibinfo {volume}
  {98}},\ \bibinfo {pages} {054406} (\bibinfo {year} {2018})}\BibitemShut
  {NoStop}%
\bibitem [{\citenamefont {H\"oll}\ \emph {et~al.}(2019)\citenamefont {H\"oll},
  \citenamefont {Kiyono},\ and\ \citenamefont {Kantz}}]{holl_PRE_2019}%
  \BibitemOpen
  \bibfield  {author} {\bibinfo {author} {\bibfnamefont {M.}~\bibnamefont
  {H\"oll}}, \bibinfo {author} {\bibfnamefont {K.}~\bibnamefont {Kiyono}}, \
  and\ \bibinfo {author} {\bibfnamefont {H.}~\bibnamefont {Kantz}},\ }\href
  {\doibase 10.1103/PhysRevE.99.033305} {\bibfield  {journal} {\bibinfo
  {journal} {Phys. Rev. E}\ }\textbf {\bibinfo {volume} {99}},\ \bibinfo
  {pages} {033305} (\bibinfo {year} {2019})}\BibitemShut {NoStop}%
\bibitem [{com({\natexlab{b}})}]{comment_2}%
  \BibitemOpen
  \href@noop {} {\emph {\bibinfo {title} {A value of $\zeta_j \approx 1$ most
  probably reflects the linear tilting trend of DWs and not their roughness.
  See Ref.~\onlinecite{holl_PRE_2019}.}}}\BibitemShut {Stop}%
\bibitem [{\citenamefont {Rosso}\ \emph {et~al.}(2003)\citenamefont {Rosso},
  \citenamefont {Hartmann},\ and\ \citenamefont {Krauth}}]{Rosso_PRE_2003}%
  \BibitemOpen
  \bibfield  {author} {\bibinfo {author} {\bibfnamefont {A.}~\bibnamefont
  {Rosso}}, \bibinfo {author} {\bibfnamefont {A.~K.}\ \bibnamefont {Hartmann}},
  \ and\ \bibinfo {author} {\bibfnamefont {W.}~\bibnamefont {Krauth}},\ }\href
  {\doibase 10.1103/PhysRevE.67.021602} {\bibfield  {journal} {\bibinfo
  {journal} {Phys. Rev. E}\ }\textbf {\bibinfo {volume} {67}},\ \bibinfo
  {pages} {021602} (\bibinfo {year} {2003})}\BibitemShut {NoStop}%
\bibitem [{\citenamefont {Kardar}\ \emph {et~al.}(1986)\citenamefont {Kardar},
  \citenamefont {Parisi},\ and\ \citenamefont
  {Zhang}}]{kardar_PRL_1986_dynamic}%
  \BibitemOpen
  \bibfield  {author} {\bibinfo {author} {\bibfnamefont {M.}~\bibnamefont
  {Kardar}}, \bibinfo {author} {\bibfnamefont {G.}~\bibnamefont {Parisi}}, \
  and\ \bibinfo {author} {\bibfnamefont {Y.-C.}\ \bibnamefont {Zhang}},\
  }\href@noop {} {\bibfield  {journal} {\bibinfo  {journal} {Physical Review
  Letters}\ }\textbf {\bibinfo {volume} {56}},\ \bibinfo {pages} {889}
  (\bibinfo {year} {1986})}\BibitemShut {NoStop}%
\bibitem [{\citenamefont {Martinez}\ \emph {et~al.}(2013)\citenamefont
  {Martinez}, \citenamefont {Emori},\ and\ \citenamefont
  {Beach}}]{martinez_APL_2013}%
  \BibitemOpen
  \bibfield  {author} {\bibinfo {author} {\bibfnamefont {E.}~\bibnamefont
  {Martinez}}, \bibinfo {author} {\bibfnamefont {S.}~\bibnamefont {Emori}}, \
  and\ \bibinfo {author} {\bibfnamefont {G.~S.~D.}\ \bibnamefont {Beach}},\
  }\href {\doibase 10.1063/1.4818723} {\bibfield  {journal} {\bibinfo
  {journal} {Applied Physics Letters}\ }\textbf {\bibinfo {volume} {103}},\
  \bibinfo {pages} {072406} (\bibinfo {year} {2013})}\BibitemShut {NoStop}%
\bibitem [{\citenamefont {Yang}\ \emph {et~al.}(2015)\citenamefont {Yang},
  \citenamefont {Ryu},\ and\ \citenamefont {Parkin}}]{yang_naturenano_2015}%
  \BibitemOpen
  \bibfield  {author} {\bibinfo {author} {\bibfnamefont {S.-H.}\ \bibnamefont
  {Yang}}, \bibinfo {author} {\bibfnamefont {K.-S.}\ \bibnamefont {Ryu}}, \
  and\ \bibinfo {author} {\bibfnamefont {S.}~\bibnamefont {Parkin}},\ }\href
  {\doibase 10.1038/nnano.2014.324} {\bibfield  {journal} {\bibinfo  {journal}
  {Nature Nanotechnology}\ }\textbf {\bibinfo {volume} {10}},\ \bibinfo {pages}
  {221} (\bibinfo {year} {2015})}\BibitemShut {NoStop}%
\end{thebibliography}%

\end{document}